\documentclass[pra,aps,twocolumn,nopacs,superscriptaddress,nofootinbib]{revtex4}

\usepackage{graphicx}
\usepackage{dcolumn}
\usepackage{bm}
\usepackage{bbm}
\usepackage{amsmath}
\usepackage{epsfig}
\usepackage{indentfirst}
\usepackage{psfrag}
\usepackage{subfigure}
\usepackage{amssymb}
\usepackage{color}
\usepackage[usenames,dvipsnames,svgnames]{xcolor}
\usepackage[colorlinks,linkcolor=blue,citecolor=blue,urlcolor=blue,hyperindex,driverfallback=dvipdfm]{hyperref}
\usepackage[T1]{fontenc}
\usepackage{comment}
\usepackage{upgreek}
\usepackage{soul}

\def\ii{{\rm i}}  \def\ee{{\rm e}}
  
\def\rb{{\bf r}}

  \def\kpar{k_\parallel}  
    
  \def\kB{{k_{\rm B}}}
\def\Eb{{\bf E}}    \def\Hb{{\bf H}}

        \def\wp{\omega_{\rm p}}
\def\epsilonm{\epsilon_{\rm m}}  
      
\def\epsilonh{\epsilon_{\rm h}}  \def\epsilonb{\epsilon_{\rm b}}  \def\wbulk{\omega_{\rm bulk}}

\begin{document}
\title{Can Copper Nanostructures Sustain High-Quality Plasmons?}

\author{Vahagn~Mkhitaryan}
\affiliation{ICFO-Institut de Ciencies Fotoniques, The Barcelona Institute of Science and Technology, 08860 Castelldefels (Barcelona), Spain}
\author{Katia~March}
\affiliation{Laboratoire de physique des solides, CNRS, Universit{\'e} Paris-Saclay, 91405 Orsay, France}
\author{Eric~Tseng}
\affiliation{Department of Materials Science and Engineering, National Taiwan University of Science and Technology, Taipei 106, Taiwan}
\author{Xiaoyan~Li}
\affiliation{Laboratoire de physique des solides, CNRS, Universit{\'e} Paris-Saclay, 91405 Orsay, France}
\author{Leonardo~Scarabelli}
\affiliation{CIC biomaGUNE, Basque Research and Technology Alliance (BRTA), Paseo de Miramón 182, 20014 Donostia-San Sebasti\'an, Spain}
\affiliation{Institut de Ciencia de Materials de Barcelona (ICMAB-CSIC), Campus de la UAB 08193 Bellaterra, Catalonia (Spain)}
\author{Luis~M.~Liz-Marz\'an}
\affiliation{CIC biomaGUNE, Basque Research and Technology Alliance (BRTA), Paseo de Miramón 182, 20014 Donostia-San Sebasti\'an, Spain}
\affiliation{Ikerbasque, Basque Foundation for Science, 38013 Bilbao, Spain}
\affiliation{Centro de Investigaci\'on Biom\'edica en Red, Bioingenier\'{\i}a, Biomateriales y Nanomedicina (CIBER-BBN), Paseo de Miram\'on 182, 28014 Donostia-San Sebasti\'an, Spain}
\author{Shih-Yun~Chen}
\affiliation{Department of Materials Science and Engineering, National Taiwan University of Science and Technology, Taipei 106, Taiwan}
\author{Luiz~H.~G.~Tizei}
\affiliation{Laboratoire de physique des solides, CNRS, Universit{\'e} Paris-Saclay, 91405 Orsay, France}
\author{Odile~St{\'e}phan}
\affiliation{Laboratoire de physique des solides, CNRS, Universit{\'e} Paris-Saclay, 91405 Orsay, France}
\author{Jenn-Ming~Song}
\affiliation{Department of Materials Science and Engineering, National Chung Hsing University, Taichung 402, Taiwan}
\author{Mathieu~Kociak}
\email{mathieu.kociak@universite-paris-saclay.fr}
\affiliation{Laboratoire de physique des solides, CNRS, Universit{\'e} Paris-Saclay, 91405 Orsay, France}
\author{F.~Javier~Garc\'{\i}a~de~Abajo}
\email{javier.garciadeabajo@nanophotonics.es}
\affiliation{ICFO-Institut de Ciencies Fotoniques, The Barcelona Institute of Science and Technology, 08860 Castelldefels (Barcelona), Spain}
\affiliation{ICREA-Instituci\'o Catalana de Recerca i Estudis Avan\c{c}ats, Passeig Llu\'{\i}s Companys 23, 08010 Barcelona, Spain}
\author{Alexandre~Gloter}
\affiliation{Laboratoire de physique des solides, CNRS, Universit{\'e} Paris-Saclay, 91405 Orsay, France}

\begin{abstract}
Silver is considered to be the king among plasmonic materials because it features low inelastic absorption in the visible and infrared (vis-IR) spectral regions compared to other metals. In contrast, copper is commonly regarded as being too lossy for plasmonic applications. Here, we experimentally demonstrate vis-IR plasmons in long copper nanowires (NWs) with quality factors that exceed a value of 60, as determined by spatially resolved, high-resolution electron energy-loss spectroscopy (EELS) measurements. We explain this counterintuitive result by the fact that plasmons in these metal wires have most of their electromagnetic energy outside the metal, and thus, they are less sensitive to inelastic losses in the material. We present an extensive set of data acquired on long silver and copper NWs of varying diameters supporting this conclusion and further allowing us to understand the relative roles played by radiative and nonradiative losses in plasmons that span a wide range of energies down to $<20\,$meV. At such small plasmon energies, thermal population of these modes becomes significant enough to enable the observation of electron energy gains associated with plasmon absorption events. Our results support the use of copper as an attractive cheap and abundant material platform for high quality plasmons in elongated nanostructures.
\end{abstract}
\date{\today}
\maketitle

\section{Introduction}

Interest in plasmons has been largely fuelled by the emergence and prospects of appealing applications, ranging from nonlinear \cite{PDN09,KZ12,SK16,JHE08} and quantum \cite{CSH06,FLK14} optics to biosensing \cite{LNL1983,WV07,KEP09,ZBH14} and light harvesting \cite{AP10,C14}. While large inelastic absorption can be beneficial for a plasmonic approach to nanoscale thermal heating \cite{B17} and cancer therapy \cite{NHH04}, dissipation reduces the ability of plasmons to locally enhance the optical electromagnetic field, which is a unique property of these excitations and the ingredient that enables several of those applications. Low dissipation and high quality factor (i.e., the dimensionless product of frequency and lifetime, $Q=\omega\tau$) are thus desirable plasmon characteristics. Among metals, silver is widely identified as an excellent choice for this reason, as visible and IR plasmons in this material are less affected by quenching originating in the coupling to electron-hole-pair transitions, which lie at higher energies. However, the search for more abundant plasmonic materials continues. In particular, aluminum has been argued to operate in the ultraviolet regime thanks to its elevated bulk plasmon frequency and still relatively low level of losses \cite{SOA13,GG15}. In this context, gold and copper, which host a conduction electron density similar to silver, are considered to be inferior for IR plasmonics because their intrinsic lifetimes ($\sim9.3$ and 6.4\,fs, respectively) are $\sim3$ and 5 times shorter than that of silver ($\sim31\,$fs) \cite{JC1972}. Still, gold is widely used because of its low reactivity and highly controlled growth \cite{GPM08,paper112}, while copper has been found to also produce well defined plasmons in nanoplates \cite{PSR09}.

Characterization of plasmons is greatly facilitated by electron energy-loss spectroscopy (EELS), which allows us to map these excitations with nanometer spatial resolution \cite{paper149,KS14,paper338}. Additionally, in contrast to far-field optical \cite{ZBH11} and cathodoluminescence \cite{paper035,VWK07,paper251} techniques, EELS is sensitive to both bright and dark modes \cite{paper121,paper251}, which is important to study plasmons that do not couple efficiently to propagating light. Visible \cite{BYT13,MKM14} and IR \cite{RB13} plasmons have been mapped using EELS with increasing degree of spectral resolution, which is now pushed to the few-meV range by taking advantage of recent advances in electron microscopy instrumentation, making it possible to explore even mid-IR optical modes, such as those associated with atomic vibrations \cite{KLD14,LTH17,HNY18,HKR19,HHP19,paper342}, as well as low-energy plasmons and their interaction with phonon polaritons \cite{paper342}.

In this Letter, we show that elongated copper nanostructures can sustain spectrally sharp vis-IR plasmons of similar quality factor as in their silver counterparts. More precisely, we use high-resolution EELS to map and spectrally characterize plasmons in both copper and silver nanowires (NWs) with lengths ranging from hundreds of nm to $>10\,\mu$m and displaying quality factors $Q>60$. We present extensive measurements on both types of NWs that allow us to fully characterize the dependence of the plasmon spectral position and width on material and geometrical parameters. The mechanism that enables copper NWs to sustain high-quality plasmons relates to the fact that a large fraction of the electromagnetic energy resides outside the metal when the light wavelength is large compared with the wire diameter, thus reducing the effect of inelastic losses. We provide electromagnetic simulations in excellent agreement with our measurements, further revealing the effect of the substrate and the relative role of nonradiative and radiative losses. Interestingly, high-quality plasmons of sub-100\,meV energy sustained by long copper NWs can also be revealed through the gain signatures that they imprint on the electron spectrum, as they are thermally populated at room temperature. These results support the use of elongated copper nanostructures for plasmonic applications requiring spectrally-narrow, long-lived modes.

\begin{figure*}
\centering
\includegraphics[width=1.0\textwidth]{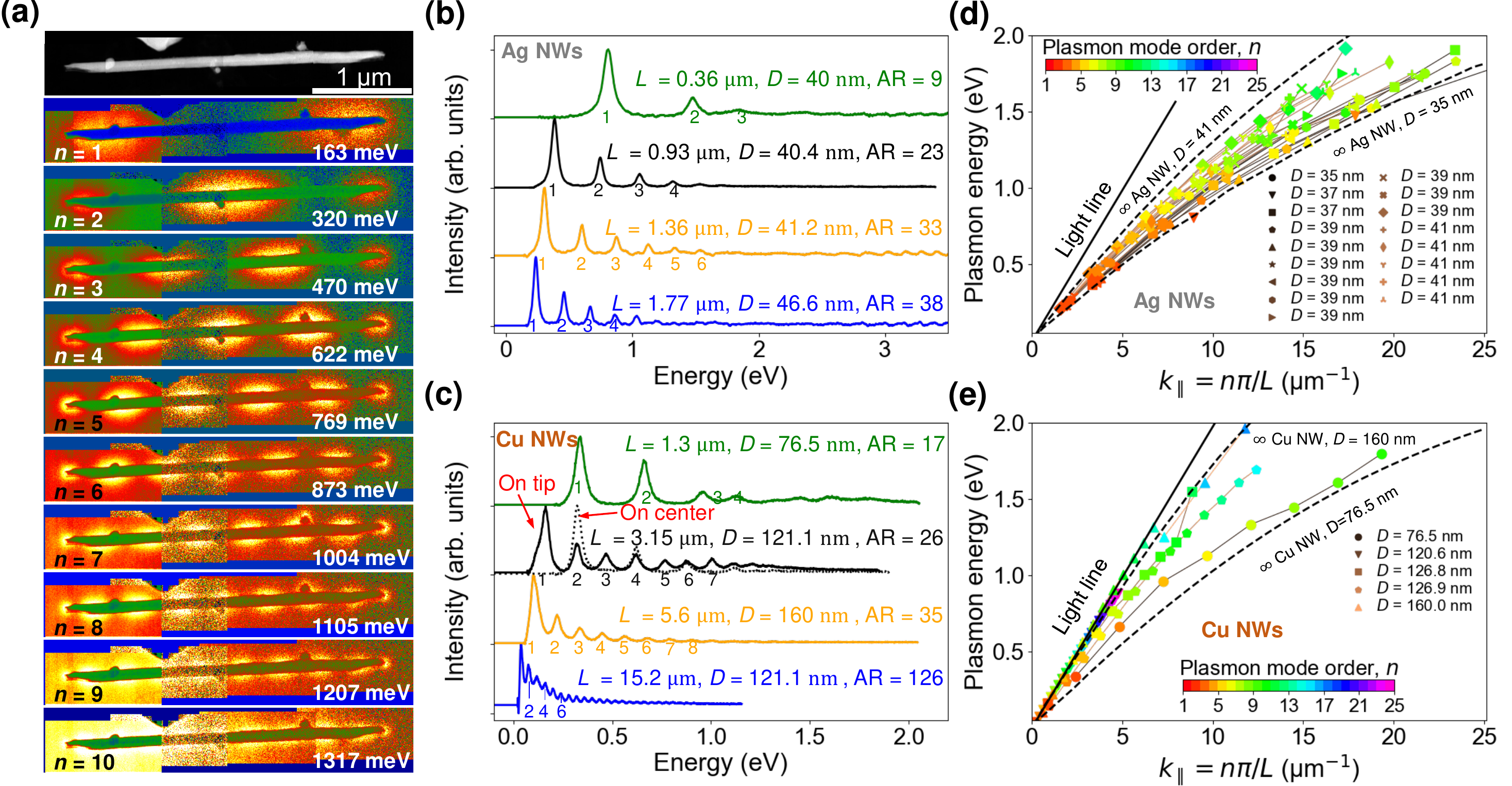}
\caption{Infrared plasmons in long Ag and Cu NWs. (a) Plasmons of increasingly high infrared energy and order $n$ sustained by a $\approx3.15$-$\mu$m-long Cu NW (see HAADF STEM image at the top) mapped through energy-filtered imaging. (b,c) Plasmonic EELS spectra collected near the tip of (b) Ag and (c) Cu NWs of varying geometrical parameters (length $L$, diameter $D$, and aspect ratio AR=$L/D$) after subtraction of the zero-loss peak (ZLP). A spectrum collected at the Cu NW center is shown in (c) (dashed curve), diplaying only modes of even order $n$, compared with excitation of all $n$'s in the spectrum acquired near the tip of the same NW (black curve). (d,e) Dispersion relations obtained from (b,c) and additional similar spectra for (d) Ag and (e) Cu NWs. We consider different NW diameters $D$ (see legends). Calculated dispersions of infinite NWs (dashed curves) are compared with data points (symbols). The latter are obtained by assigning a parallel wave vector $\kpar=n\pi/L$ depending on mode order $n$ (see scales with symbol color) for each NW length $L$. NWs are deposited on a 15-nm-thick Si$_3$N$_4$ substrate.}
\label{Fig1}
\end{figure*}


We present in Figure\ \ref{Fig1} a comprehensive analysis of plasmons in Ag and Cu NWs for a wide range of geometrical parameters, resulting in mode energies that cover a range extending from $\sim20\,$meV to the onset of interband transitions in these metals. We base our analysis on state-of-the-art EELS with a combined energy and space resolution of $\sim10\,$meV and $\sim1\,$nm, respectively, as described in Methods. Figure\ \ref{Fig1}a shows energy-filtered EELS maps of a Cu NW of $\approx3.15\,\mu$m length, exhibiting  a number of maxima given by $n+1$, where $n$ labels the order of the plasmon mode (i.e., this is the number of nodes in the plasmonic standing waves along the NW \cite{N07_2,RCV11,NWM11,PM14,KS14,MKM14,paper258}). These characteristic standing waves have clear associated spectral signatures, as depicted in Figure\ \ref{Fig1}b for Ag and in Figure\ \ref{Fig1}c for Cu NWs. Both Ag and Cu spectra manifest the expected quasi-harmonic series of peaks, shifting to lower energy as the aspect ratio (AR$=L/D$) increases and following the dispersion relation defined for wires of infinite length, as represented in Figure\ \ref{Fig1}d,e. However, closer and more quantitative analysis reveals major differences between both types of NWs. A first observation is an expected dependence of the dispersion relation on wire diameter, with thinner NWs deviating more from the light line. Additionally, due to their longer lengths, Cu NWs diplay plasmons at lower energies. More strikingly, we clearly show that the dispersion relations for the Cu plasmons are much closer to the light line than those of silver NWs (i.e., Cu plasmons show a more retarded behavior than Ag plasmons); specifically, Cu NWs with lengths above $5\,\mu$m and diameters of the order of $160\,$nm permit reconstructing dispersion curves nearly indistinguishable from the light line up to plasmon energies of 1\,eV. Deviations occur for shorter NW lengths or smaller diameters. Incidentally, in Figure\ \ref{Fig1}e, a series of plasmon modes with a substantial departure from the light line (in a behavior similar to Ag NWs) is observed for a Cu NW with a length of 1.3\,$\mu$m and a relatively small diameter $D=76.5$\,nm.

\begin{figure*}
\centering
\includegraphics[width=0.8\textwidth]{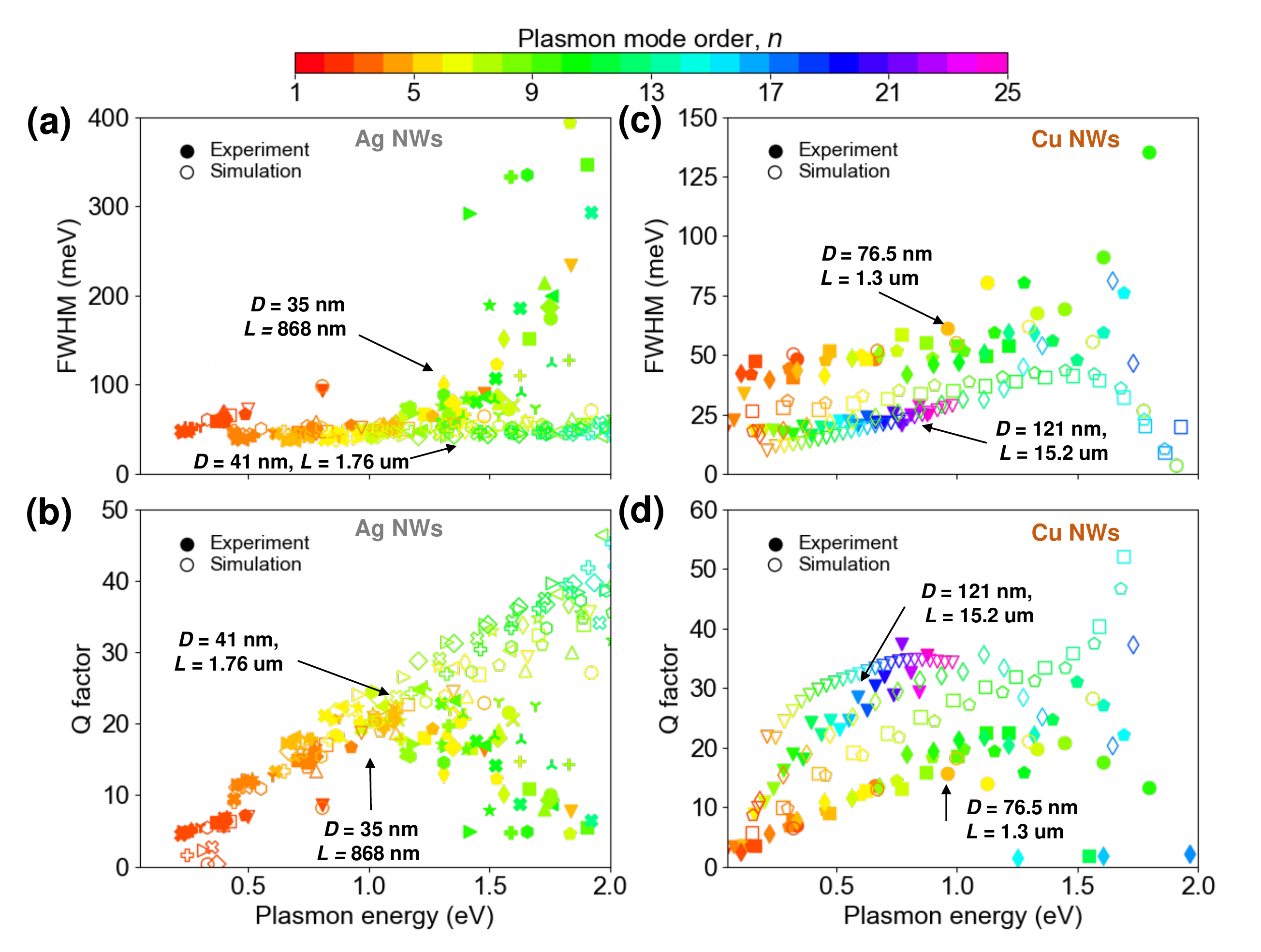}
\caption{Plasmon width and quality factor. We show experimental (filled  symbols) and simulated (open  symbols) values of the FWHM (a,c) and quality factor $Q$ (b,d) for Ag (a,b) and Cu (c,d) NWs collected over a broad range of geometrical parameters and mode order $n$. The FWHM is obtained by Lorentzian fitting. NW diameters are encoded in the symbol types, which are the same as in Figure\ \ref{Fig1}d,e, with maximum and minimum values as indicated by labels in each plot.}
\label{Fig2}
\end{figure*}

Major differences between the two metals also occur in the  behavior of the FWHM and quality factor (see Methods), which we plot as a function of plasmon energy for Cu and Ag NWs in Figure\ \ref{Fig2}. For Ag NWs, the experimentally observed FWHMs follow a monotonic increase from 40\,meV to 400\,meV as the plasmon energy is raised. At low energy, the FWHM of dipolar modes ($n=1$, red marks) display a very slow increase with energy. A similar behavior is encountered in other low-$n$ modes. However, for each value of the plasmon energy, the FWHM decreases with mode order $n$ in this regime. The FWHMs of plasmons at all orders continue to slowly increase with resonance energy until this approaches $\approx$1.8\,eV. Around this energy, the FWHM suddenly increases dramatically from less than 100\,meV to values exceeding 400\,meV. We attribute this behavior to the presence of a Au seed core used in the growth of Ag NWs \cite{paper258} (see Figure\ \ref{FigS5} and also below).

The FWHM in the studied Cu NWs is systematically smaller than in Ag NWs. In Figure\ \ref{Fig2}c, the FWHM of Cu NWs varies from 20\,meV to 150\,meV. As a general trend, the widths increase with plasmon energy, which is a behavior also observed in Ag NWs. Additionally, the minimum energy width is actually not found in the dipolar mode, but often in multipolar modes at resonance energies of $\sim300\,$meV. We also analyze much longer ($>15\,\mu$m) nanowires, for which the FWHM is smaller than the instrumental spectral resolution. A near constant FWHM at low energies translates into a linear increase in quality factor $Q$ with plasmon energy (Figure\ \ref{Fig2}b,d). For Ag NWs (Figure\ \ref{Fig2}b), the quality factors steadily increase from $\sim4$ to $\sim25$ and start decreasing above resonance energies around 1.3\,eV. In Cu NWs (Figure\ \ref{Fig2}d), even larger quality factors are observed (e.g., $Q\sim25$ and 35 in Ag and Cu NWs at 1\,eV, respectively).

Incidentally, Ag nanowires are grown starting from a $\sim200\,$nm Au core \cite{paper258}, much smaller than the wire lengths under consideration, but still capable of producing inelastic losses at the onset of interband transitions above $\sim1.5-2.5\,$eV in Au (compared with $\sim4\,$eV in Ag). We attribute part of the onset of losses in the Ag NWs above that energy (Figure\ \ref{Fig2}a,b) to losses in the gold core. In contrast, simulations for pure Ag NWs (i.e., without Au core) are immune to this effect. This conclusion is consistent with numerical simulations performed with and without inclusion of the Au core (see Figure\ \ref{FigS5}), where the former undergo a sudden increase in plasmon width above 2\,eV.

\begin{figure*}
\centering
\includegraphics[width=0.6\textwidth]{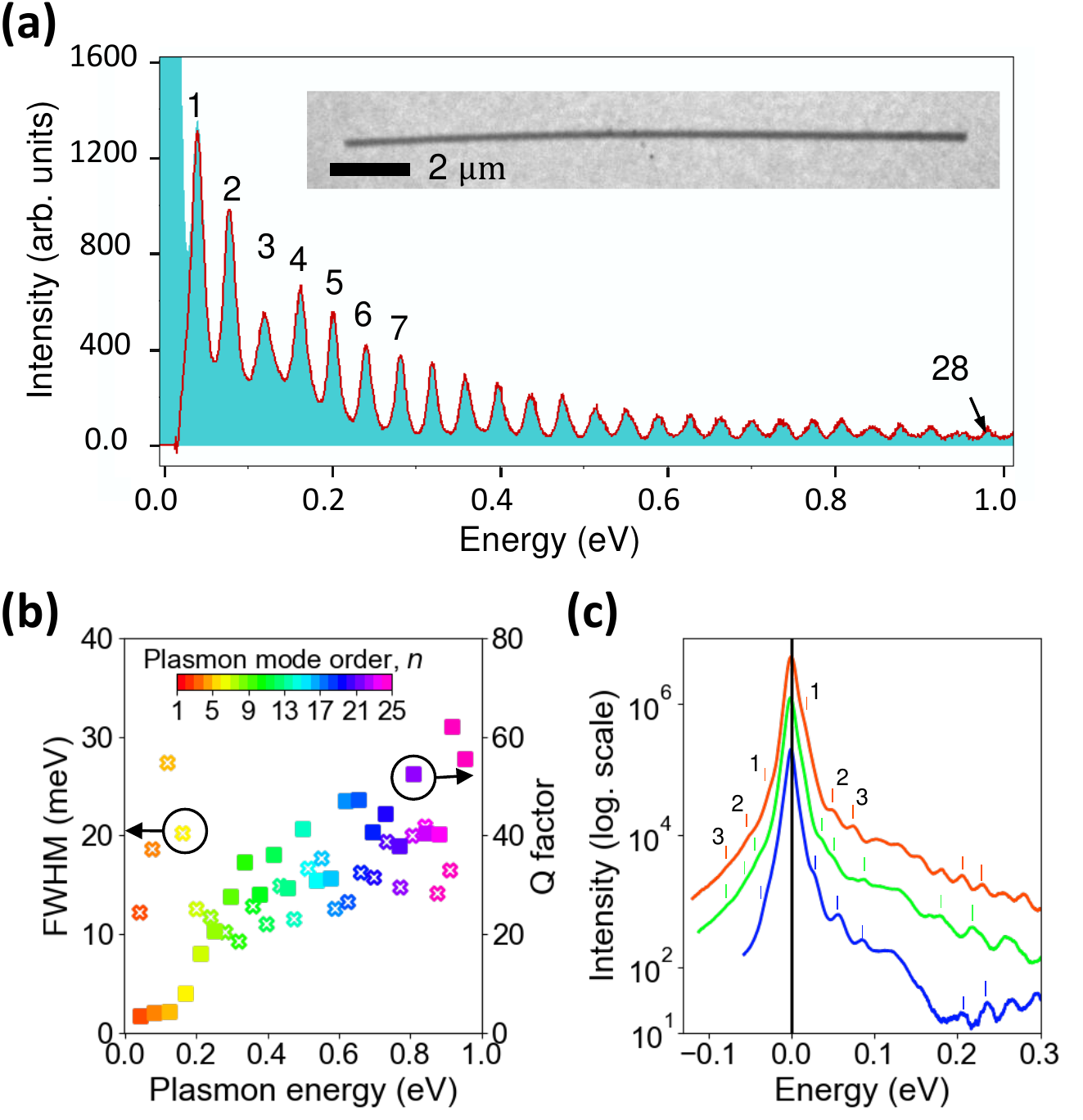}
\caption{Electron-induced absorption and emission of infrared plasmons in a long Cu NW. (a) EELS spectrum measured at the tip of a long Cu NW ($15.2\,\mu$m length, $121\,$nm diameter, see bright-field image in the inset). (b) FWHM (crosses, left scale) and quality factor $Q$ (squares, right scale) obtained from the spectrum in (a). (c) EELS spectra revealing plasmonic energy gains associated with thermally excited IR plasmons, as probed by the electron beam focused near the tip of three $L>20\,\mu$m Cu NWs.}
\label{Fig3}
\end{figure*}

Compared to the results shown in Figures\ \ref{Fig1} and\ \ref{Fig2}, the spectra of longer nanowires need to be deconvoluted from the ZLP in order to increase the spectral resolution, which we bring from 14\,meV to 8\,meV by using the Richardson-Lucy algorithm \cite{GDT03} (see Section\ \ref{secS2} and Figures\ \ref{FigS2} and \ref{FigS3}). We present an example of this procedure in Figure\ \ref{Fig3}a, showing a spectrum acquired near the tip of a long Cu NW ($15.2\,\mu$m length, 121\,nm diameter) and characterized by a long series of plasmon standing waves that reach an order $n=28$. FWHM analysis of this spectrum (Figure\ \ref{Fig3}b) reveals values down to 10\,meV in plasmons of $\sim300\,$meV energy, whereas the quality factor exceeds $Q>60$ at $\sim1\,$eV plasmon energy. We note that these long Cu NWs sustain well-defined plasmons at energies that are low enough to be thermally activated at room temperature (i.e., for $\hbar\omega\lesssim\kB T\sim26\,$meV), as emphasized by the presence of energy gain peaks observed in the measured spectra of Figure\ \ref{Fig3}c, thus extending the field of EELS thermometry from optical phonons \cite{LB18,ILT18} to mid-IR plasmons. The ratio of loss-to-gain peak intensities is determined from the Bose-Einstein distribution function $n_\omega$ as $(n_\omega+1)/n_\omega=\ee^{\hbar\omega/\kB T}$, which produces results compatible with a temperature $T\sim300\,$K upon detailed analysis of the data in Figure\ \ref{Fig3}c, thus corroborating the thermal origin of the observed plasmon gains. Incidentally, the explored energy range includes phonon losses from the substrate, particularly at higher energies $\gtrsim100\,$meV \cite{L99_2,KMT08}, and we note that phonon-plasmon hybridization can reduce $Q$, as observed in silver NWs coupled to boron nitride phonons \cite{paper342}.

\begin{figure*}
\centering
\includegraphics[width=0.7\textwidth]{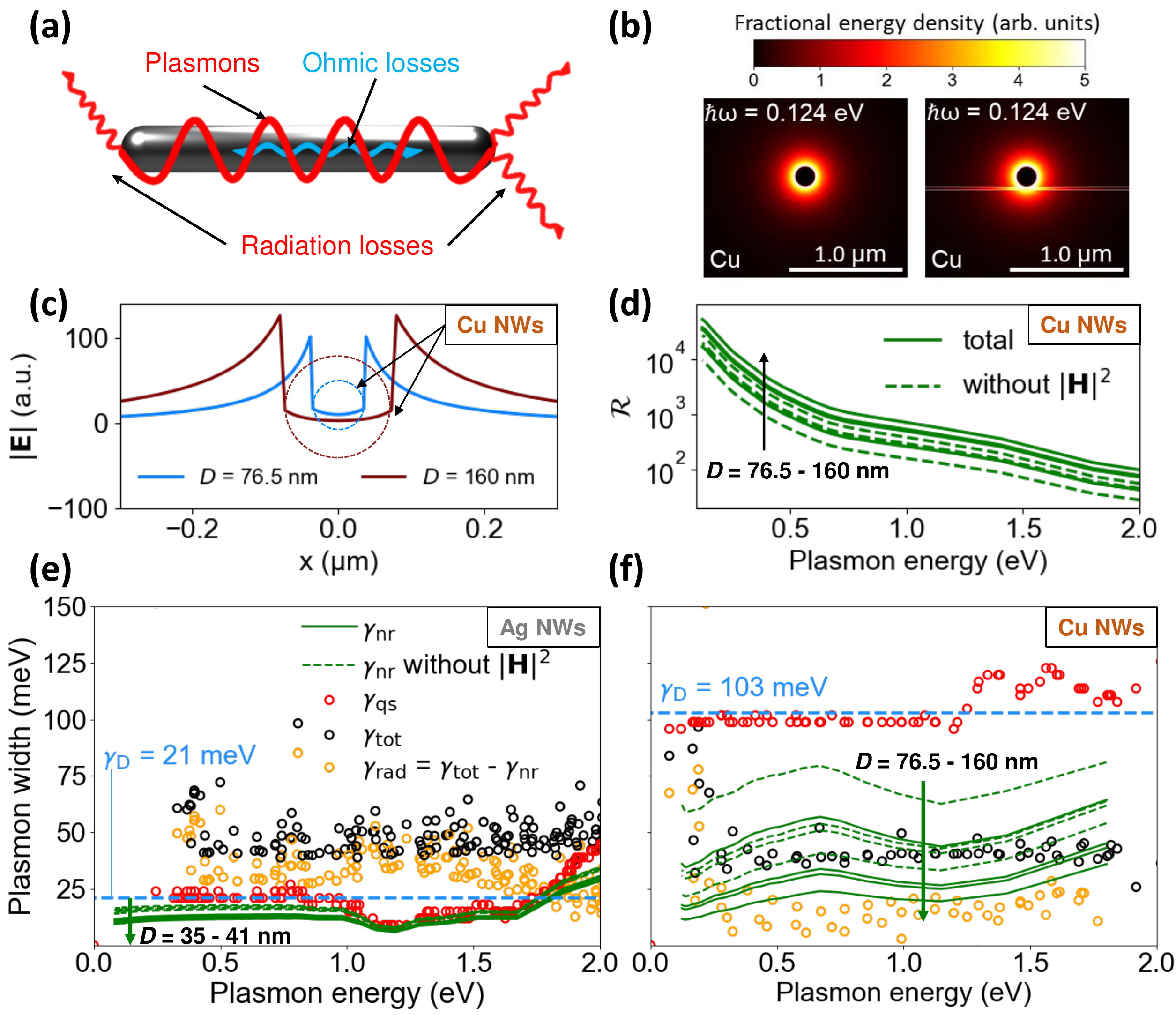}
\caption{Radiative and nonradiative losses in the plasmons of Cu and Ag NWs. (a) Schematic representation of these loss mechanisms. (b) Calculated cross section of the electromagnetic energy density distribution in self-standing (left) and supported (right, 15-nm-thick Si$_3$N$_4$ substrate) Cu NWs of 76.5\,nm diameter and infinite length at 124\,meV plasmon energy. (c) Radial distribution of the plasmonic electric near-field amplitude in self-standing infinite Cu NWs of diameters 76.5\,nm and 160\,nm (indicated by dashed circles) for 124\,meV plasmon energy. (d) Ratio of the spatially integrated electromagnetic energy outside and inside self-standing Cu NWs of different diameters ($\mathcal{R}$, eq\ \ref{Rterm}). Solid and dashed curves are calculated with and without inclusion of the magnetic field contribution. (e,f) Plasmon width contributions in (e) Ag and (f) Cu NWs: green curves represent the nonradiative rate $\gamma_{\rm nr}$ calculated from eq\ \ref{gammanr} with (solid curves) and without (broken curves) inclusion of the magnetic field for a range of increasing diameters $D$ as indicated by the green arrows; black symbols show the total width $\gamma_{\rm tot}$ extracted from the FWHM simulations presented in Figure\ \ref{Fig2}a,c for NWs within the same range of diameters and different lengths; orange symbols correspond to the radiative damping $\gamma_{\rm rad}$ estimated by subtracting $\gamma_{\rm nr}$ from the black circles; the Drude damping (blue) and the quasistatic damping ($\gamma_{\rm qs}$, red) are shown for comparison.}
\label{Fig4}
\end{figure*}

Further understanding of plasmons in Cu and Ag NWs can be gained by examining radiative and nonradiative loss mechanisms from a theoretical viewpoint. We first note that the observed plasmons display symmetric patterns along the transverse wire direction (see Figure\ \ref{Fig1}a), and in fact they correspond to modes that are axially symmetric in self-standing structures, with a small deformation from axial symmetry being produced by the presence of a thin substrate (see Figure\ \ref{Fig4}b). Chiral plasmons with an $\ee^{\ii m\varphi}$ azimuthal dependence and $m\neq0$ are also supported, but they emerge at higher energies and leave a comparatively weak trace on the transmitted electrons \cite{paper116}. The dominant NW plasmons has $m=0$ symmetry and can be understood as Fabry-Perot (FP) resonances in a cavity, whereby the round-trip propagation along the length $L$ is accounted for by a phase $2\kpar L$, where $\kpar$ is the material-, diameter-, and frequency-dependent wave vector in an infinite wire (Figure\ \ref{Fig1}d,e), while reflection at the ends can be described through a coefficient $r$. The FP condition that signals the existence of plasmons becomes $\kpar L+{\rm arg}\{r\}=n\pi$, where $n$ is the mode order, which we use as a label in Figures\ \ref{Fig1}-\ref{Fig3}. The presence of antinodes at the NW ends indicates that the phase of the so-defined reflection coefficient is ${\rm arg}\{r\}\sim0$ (therefore, we ascribe \cite{N07_2} $\kpar=n\pi/L$ in Figure\ \ref{Fig1}d,e), whereas the high quality factors observed from the EELS plasmon features reveal that they propagate multiple times along the NW length during their lifetimes, and therefore $|r|\lesssim1$. With this perspective in mind, we can consider that the plasmon modes are bound to the NW during propagation, while scattering at the ends can result in out-coupling to radiation, as illustrated in Figure\ \ref{Fig4}a.

The quality factor of plasmons in Ag and Cu structures depends on the contribution to the dissipation rate coming from both radiative and nonradiative mechanisms. In particular, the nonradiative contribution is given by the ratio of the absorption power to the electromagnetic energy $\gamma_{\rm nr}=P_{\rm abs}/W$. These quantities are in turn related to the monochromatic electric and magnetic field intensities at the mode frequency $\omega$ combined with the position- and frequency-dependent local permittivity $\epsilon(\rb,\omega)$ as \cite{J99}
\begin{align}
&P_{\rm abs}(\omega)=\frac{\omega}{2\pi}\int d^3\rb\,|\Eb(\rb,\omega)|^2\,{\rm Im}\{\epsilon(\rb,\omega)\},\nonumber\\
&W(\omega)=\frac{1}{4\pi}\int d^3\rb\,\big[|\Eb(\rb,\omega)|^2\,\partial_\omega{\rm Re}\{\omega\,\epsilon(\rb,\omega)\} \nonumber\\
&\quad\quad\quad\quad\quad\quad\quad\;+|\Hb(\rb,\omega)|^2\big].
\nonumber
\end{align}
Neglecting absorption at the NW ends and treating the plasmon mode in the above FP model, we need to evaluate these expressions for the $m=0$ guided mode of an infinite wire as a function plasmon frequency $\omega$. We find a nonradiative contribution to the decay rate
\begin{align}
\gamma_{\rm nr}=\frac{P_{\rm abs}}{W}=\frac{2\omega\,{\rm Im}\{\epsilonm(\omega)\}}{\partial_\omega{\rm Re}\{\omega\,\epsilonm(\omega)\}+\mathcal{R}(\omega)},
\label{gammanr}
\end{align}
where
\begin{align}
\mathcal{R}(\omega)=\dfrac{\int d^3\rb\,|\Hb(\rb,\omega)|^2+\int_{\rm out} d^3\rb\,\epsilonh(\rb)\,|\Eb(\rb,\omega)|^2}{\int_{\rm in} d^3\rb\,|\Eb(\rb,\omega)|^2},
\label{Rterm}
\end{align}
the {\it in} and {\it out} integration domains refer to the volume inside and outside the metal, respectively, $\epsilonm(\omega)$ is the metal permittivity, and the host permittivity is $\epsilonh(\rb)=1$ for self-standing NWs, but it depends on position (inside or outside the substrate) for supported NWs.

In the quasistatic (qs) limit, which is valid for small structures compared to the light wavelength, the spectral line shape associated with the excitation of a plasmon of frequency $\wp$ reduces to \cite{paper300} $\sim{\rm Im}\{1/[\epsilonm(\omega)-\epsilonm(\wp)]\}$ both in EELS and in the optical extinction. In this limit, $\wp$ depends on material and morphology, but not on the size of the structure. Additionally, radiative losses vanish, so the total decay rate $\gamma_{\rm qs}$ is given by the FWHM of the above expression. In particular, for a Drude-like response $\epsilonm(\omega)=\epsilonb-\wbulk^2/\omega(\omega+\ii\gamma)$, which provides a good fit to the measured permittivity of Ag ($\epsilonb=4.0$, $\hbar\wbulk=9.17\,$eV, $\hbar\gamma=21\,$meV) and Cu ($\epsilonb=8.0$, $\hbar\wbulk=8.88\,$eV, $\hbar\gamma=103\,$meV) at frequency $\omega$ below their respective interband regions \cite{JC1972,paper300}, the quasistatic line shape becomes $\sim{\rm Im}\{1/[\wp^2-\omega(\omega+\ii\gamma)]\}$ and the decay rate is $\gamma_{qs}\approx\gamma$ for large plasmon energy compared with the intrinsic damping rate, $\wp\gg\gamma$. Reassuringly, inspection of eq\ \ref{gammanr} further reveals $\gamma_{\rm nr}\approx\gamma$ in the quasistatic Drude approximation.

Figure\ \ref{Fig4}c shows the simulated field amplitude distribution across the diameter of self-standing Cu NWs with diameters $D=76.5$ and 160\,nm and infinite length, clearly showing a larger weight outside the metal, and thus anticipating a reduction of nonradiative losses relative to the quasistatic limit. We corroborate this conclusion by examining the ratio $\mathcal{R}$ (eq\ \ref{Rterm}) in Figure\ \ref{Fig4}d, where we find $\mathcal{R}\gg1$ (the effect is dramatic at low energy as a result of the larger delocalization of the field, see Figure\ \ref{Fig4}b), although the outside contribution to the electromagnetic energy decreases with plasmon frequency as the NW dispersion relation evolves away from the light cone (Figure\ \ref{Fig1}d,e). Additionally, the presence of a substrate contributes to increase localization (Figure\ \ref{Fig4}b), thus lowering $\mathcal{R}$ at low energy.

The results of this theoretical analysis are presented in Figure\ \ref{Fig4}e,f for Ag and Cu NWs within the range of diameters and lengths experimentally explored in this work. We compare the total plasmon width $\gamma_{\rm tot}$ (black symbols, extracted from the calculations in Figure\ \ref{Fig2}a,c) to the nonradiative ($\gamma_{\rm nr}$, eq\ \ref{gammanr}, green curves) and radiative ($\gamma_{\rm rad}=\gamma_{\rm tot}-\gamma_{\rm nr}$, orange symbols) contributions. As a general conclusion, we find that Ag NWs show a total damping that is slightly larger than Cu NWs, a result that we attribute to an increase of radiative losses in the former due to their smaller diameter: although nonradiative losses in Cu are higher (cf. blue curves and red symbols in Figure\ \ref{Fig4}e,f), the larger diameter of the Cu NWs results in substantially smaller radiative damping. Additionally, the ratio $\mathcal{R}$ in eq\ \ref{Rterm} increases with diameter (see Figure\ \ref{Fig4}d), thus contributing to lower nonradiative damping as well (eq\ \ref{gammanr}). We note however that, although $\mathcal{R}$ is higher at lower energies, $\epsilonm(\omega)$ also increases in this limit (e.g., in the Drude limit, $\omega\,{\rm Im}\{\epsilonm(\omega)\}$ and $\partial_\omega{\rm Re}\{\omega\,\epsilonm(\omega)\}$ scale as $\gamma\wbulk^2/(\omega^2+\gamma^2)$ and $\epsilonb+\wbulk^2(\omega^2-\gamma^2)/(\omega^2+\gamma^2)^2$, respectively), so the overall dependence of $\gamma_{\rm nr}$ on frequency results from the interplay between these two quantities, and in fact, we observe in Figure\ \ref{Fig4}e,f (green curves) a rather featureless nonradiative damping. We also note that $\gamma_{\rm rad}$ depends on how good the coupling to the far field is, which in turn is highly sensitive to NW geometry and mode order \cite{MKM14}. In particular, low-order modes have stronger coupling to the far field. Importantly, the presence of the Si$_3$N$_4$ substrate is found from theory to lower the quality factor of Ag NWs by $\sim30\%$ relative to self-standing NWs (see Figure\ \ref{FigS5}). Incidentally, in much longer NWs (e.g., $\sim100\,\mu$m in Figure\ \ref{FigS4}), the spectral resolution of the microscope is not enough to discern low-energy plasmons close to the ZLP, but high-order modes can be identified with an energy spacing as low as 10-15\,meV thanks to the achieved 8-10\,meV EELS resolution. 

\section{Conclusions}

In conclusion, by systematically investigating high-resolution EELS of micron-sized silver and copper NWs, we have unveiled high-quality vis-IR plasmons in both of these systems, thus challenging the commonly accepted view that copper is an inferior plasmonic material. We show that a substantial fraction of the electromagnetic energy associated with plasmons resides outside the metal, thus explaining the reduction in the relative effect of ohmic losses in these modes. The interplay between radiative and nonradiative losses results in measured quality factors as high as $Q>60$ in the visible part of the spectrum. Our measurements extend up to very low plasmon energies ($<20\,$meV) in the mid-IR, where thermal population at room temperature is enough to produce traces in the energy gain side of the electron spectra. The observed mechanism of loss reduction is promising for further applications of plasmons because it enables high quality modes even in intrinsically lossy materials such as Cu, which could offer additional functionalities related to their thermal, electrical, and thermoelectrical properties. Other cheap and abundant metals could be envisaged, provided they can be grown in the form of long and thick structures, as shown here for copper nanowires.

\appendix
\label{Methods}

\section{NW synthesis and sample preparation}

Cu NWs were synthesized following a thermally assisted photoreduction process \cite{TSD08} (see Figure\ \ref{FigS1} and mode synthesis details below). Ag NWs were synthesized following a previously reported method \cite{paper258} for controlled growth of the wire length with nearly constant diameter. NW lengths in the sub-$\mu$m to 10s\,$\mu$m range were produced using this method. Solutions of Cu and Ag NWs were then drop-casted on separate 15-nm-thick ${\rm Si_3N_4}$ membranes (Ted Pella 21569-10) for TEM analysis.

Copper nanowires (NWs) were synthesized following a thermally assisted photoreduction process \cite{TSD08}. This starts by producing substrates consisting of TiO${_2}$ thin films deposited on a Si wafer by coating a gel solution. The gel was in turn prepared using isopropyl alcohol (J.T. Baker)/titanium(IV) isopropoxide (Sigma Aldrich, 97\%)/hydrochloric acid (J.T. Baker, 36.5-38.0\%) with a volume ratio of 170:12:0.4 and stirred for 10 min before aging at room temperature for 48 h. The as-synthesized TiO${_2}$ thin films were annealed at 500$^{\circ}$C for 8\,h in oxygen atmosphere to achieve a well crystallized anatase TiO${_2}$. The Cu NWs were dispersed in 0.1\,M copper (II) chloride (Aldrich, 99.999\%) aqueous solution, and 10\,$\mu$l droplets of this solution were dropped on the TiO${_2}$-coated Si substrates, followed by isothermal heating at 400$^{\circ}$C for 3\,h in 97\% N${_2}$-3\% H${_2}$ inside an infrared (IR) furnace, and subsequent furnace-cooling back to ambient temperature. Crystallographic and compositional details of the so obtained Cu NWs are shown in Figure\ \ref{FigS1}.

\section{Electron microscopy and EELS}

Samples were studied in two NION monochromated HERMES machines operated at 60\,kV (for Cu and Ag NWs) and 30\,kV (Cu NWs) acceleration voltages. The 60\,kV NION at Arizona State University was fitted with a Gatan Quantum spectrometer and operated at a dispersion of 2 meV per channel with incidence and acceptance semi-angles of 12\,mrad and 14\,mrad, respectively; the typical ZLP FWHM was 16-20\,meV; the individual spectrum acquisition time was determined by the requirement of recording the ZLP close to saturation (typically hundreds of ms); spectra were corrected from gain. The NION electron microscope at University of Paris-Saclay (30 and 60\,kV) was fitted with a NION IRIS monochromator and a Kuros camera; the incidence and acceptance semi-angles were 10-15\,mrad and $<15\,$mrad, respectively; the energy dispersion was 0.2\,meV/channel for the highest-resolved EELS (ZLP FWHM of $\approx10\,$meV), and 1-3\,meV/channel for a ZLP FWHM of $\approx20\,$meV; spectra were obtained through time-series acquisition (i.e., several hundreds of spectra were typically acquired with individual collection times in the 1-100\,ms range, and subsequently realigned to compensate tiny energy shifts and residual 50\,Hz shifts in the shortest acquisition times). All spectra were deconvoluted to subtract the ZLP as explained in Section S2 of SI \cite{GDT03}. Experimental plasmon widths were obtained by fitting EELS spectra to Lorentzians \cite{BYT13,V14} and compared with a similar fitting of calculated spectra.

\noindent {\bf Electromagnetic simulations.} Numerical simulations of supported NWs were performed using the Lumerical FDTD package. We compared experimentally measured EELS spectra with the calculated local-density of optical states (LDOS). The latter was obtained from the induced field acting on a source dipole \cite{paper102} as a function of photon frequency and position for a dipole oriented along the electron beam direction. EELS probabilities and the LDOS were also obtained for self-standing NWs using the boundary-element method \cite{paper040}. Plasmon dispersion relations of infinite NWs (Figure\ \ref{Fig1}d,e) were obtained analytically \cite{AE1974}. Measured dielectric functions of Cu and Ag metals \cite{JC1972} and Si$_3$N$_4$ insulator \cite{P1985} were used in the simulations.

\section{Deconvolution of EELS instrumental broadening and retrieval of plasmon full width at half maximum (FHWM)}
\label{secS2}

We first evaluated the FWHM of the observed plasmon features through Lorentzian fits, which commonly yield good results in EELS \cite{BYT13}. When the obtained FWHM was larger than the zero-loss peak (ZLP) by more than a factor of $\sim2$, the ZLP contribution to the peak broadening was less than $\approx12\%$ and could be neglected. For simplicity, the FHWM and Q factors shown in Figure\ 2 were obtained using Lorentzian fitting. However, we could measure some Lorentzian FWHMs as small as 15\,meV in long Cu NWs ($\geq10\,\mu$m), to be compared with a ZLP FWHM of $\sim10\,$meV. The presented FWHM and quality factors in Figure\ 2 were then respectively over- and under-estimated, and they could be understood as a conservative estimate of the quality of Cu NW resonators. In order to obtain a more accurate estimate of the FWHM and quality factor $Q$ of long Cu NWs, we implemented a deconvolution of instrumental broadening of the EELS spectra. We used this method to obtain the deconvoluted spectrum shown in Figure\ 3a for a 15.2\,$\mu$m copper NW acquired with the electron beam positioned close to one of the NW tips. The energy resolution was increased from 14 to 8\,meV by deconvolution of the ZLP based on the Richardson-Lucy algorithm \cite{GDT03} (see also Figures\ \ref{FigS2} and \ref{FigS3}).

Next, we explain the methods that we use to remove the ZLP contribution. In Figure\ \ref{FigS2}, we analyze a NW hosting a plasmon at an energy of 25\,meV. The EELS raw data are plotted in log scale in Figure\ \ref{FigS2}a as obtained using 30\,keV electrons with an energy resolution of 11\,meV estimated from the FWHM of the ZLP. The dipolar mode ($n=1$) produces a faint shoulder on the ZLP side, while modes $n>1$ give rise to clear peaks in the spectrum. The energy resolution of the EELS data is increased by Richardson-Lucy deconvolution with 10 iterations \cite{GDT03} and the obtained spectrum has now an energy resolution of 6.5\,meV, so plasmonic peaks become sharper (Figure\ \ref{FigS2}a, right). Both raw and deconvoluted data show evidence of broader modes $n=4$-6 (at 100-160\,meV energy loss).
	
In order to evaluate the influence of instrumental broadening, fitting with pseudo-Voigt functions, deconvolution, or a combination of deconvolution and pseudo-Voigt-function fitting are compared with the results from a pure Lorentzian fitting.	Figure\ \ref{FigS2}b shows the results for the raw data fitted with Lorentzians or with pseudo-Voigt functions, as well as the deconvoluted spectra fitted with pseudo-Voigt functions. Taking the example of the peak for mode $n=3$ at an energy of 74\,meV, Lorentzian fitting gives a FWHM of 22.1\,meV, whereas fitting with a pseudo-Voigt function gives a FWHM of 20.0\,meV (a similar result is obtained by fitting with a Voigt function). The pseudo-Voigt width takes into account the natural Lorentzian broadening of the plasmon peak and a Gaussian broadening due the instrumental response, with the latter estimated at 11\,meV. Using an approximation detailed in Ref. \citenum{V14}, it is possible to retrieve a Lorentzian broadening of 13.7\,meV for the $n=3$ mode. Following a similar approach for the spectra after Richardson-Lucy deconvolution, a peudo-Voigt broadening of 15.0\,meV is obtained, corresponding to a natural Lorentzian broadening of 12.1\,meV after removing the 6.5\,meV Gaussian contribution. We note that both approaches (Voigt fitting of raw data or after ZLP deconvolution) yield rather similar final natural broadening (i.e., 12-14\,meV), while the instrumental broadening is substantial before deconvolution (11\,meV) and produces just a small correction after deconvolution (6.5\,meV).  

A similar correction procedure confirms the strong broadening of the $n=4$ mode at 103\,meV with a Lorentzian broadening of 23-25 meV. Additionally, the Lorentzian broadening is estimated to be as low as 10-12\,meV for the plasmon at 200\,meV energy ($n=8$). However, the broadening of the dipolar mode ($n=1$) is more difficult to retrieve because it lies on top of the ZLP tail, so we do not discuss it for long Cu NWs. Figure\ \ref{FigS2}c shows a comparison of the FWHM and the quality factors obtained from a pure Lorentzian fit when instrumental broadening has been corrected. The obtained widths correspond to modes ranging from $n=2$ (52\,meV energy) to $n=13$ (330\,meV energy).

Figure\ \ref{FigS3} shows a similar treatment for EELS data acquired from the tip of a Cu NW and showing a plasmon feature at an energy of $\sim38\,$meV. Figure\ \ref{FigS3}a portrays the raw EELS data obtained from 60\,keV electrons with an energy resolution of 14\,meV (left), along with the spectrum found after deconvolution (right), which results in an energy resolution of 8\,meV. The corresponding FWHM and quality factor are shown in Figure\ \ref{FigS3}b. We conclude by discussing the results obtained after instrumental deconvolution (Lorentzian contribution after RL deconvolution). The FWHM for the dipolar mode is $\sim11$-$12\,$meV and could be slightly underestimated because it is affected by the ZLP subtraction that tends to decrease its value. The FWHMs for the modes $n=2$-4 are larger, in the 20-30\,meV range, presumably due to stronger coupling to the Si$_3$N$_4$ substrate. The FWHM then decreases to reach a minimum of $\sim9$-10\,meV at 300\,meV plasmon energy (mode $n=8$) and then increases with plasmon energy to reach $\sim20\,$meV for $n=21$ (800\,meV plasmon energy). The corresponding quality factors are 20 and 40 at 250\,meV and 800\,meV energy, respectively.

\section*{Acknowledgments}

We acknowledge the use of (S)TEM at the John M. Cowley Center for High Resolution Electron Microscopy in the Eyring Materials Center at Arizona State University. This work has been supported in part by the National Agency for Research under the program of future investment TEMPOS CHROMATEM (Ref. No. ANR-10-EQPX-50), the 2017-2018 France-Taiwan Orchid Program with (Ref. No. 106-2911-I-005-501), the European Commission (Grant No. 823717 ESTEEM3), the European Research Council (Advanced Grant 789104-eNANO), the Spanish MINECO (MAT2017-88492-R, MAT2017-86659-R, and SEV2015-0522), the Catalan CERCA Program, and Fundaci\'o Privada Cellex. L.S. acknowledges support from the Marie Sklodowska‐Curie Actions SHINE (H2020-MSCA-IF-2019, Grant No. 894847).


\begin{figure*}
\centering
\includegraphics[width=0.75\textwidth]{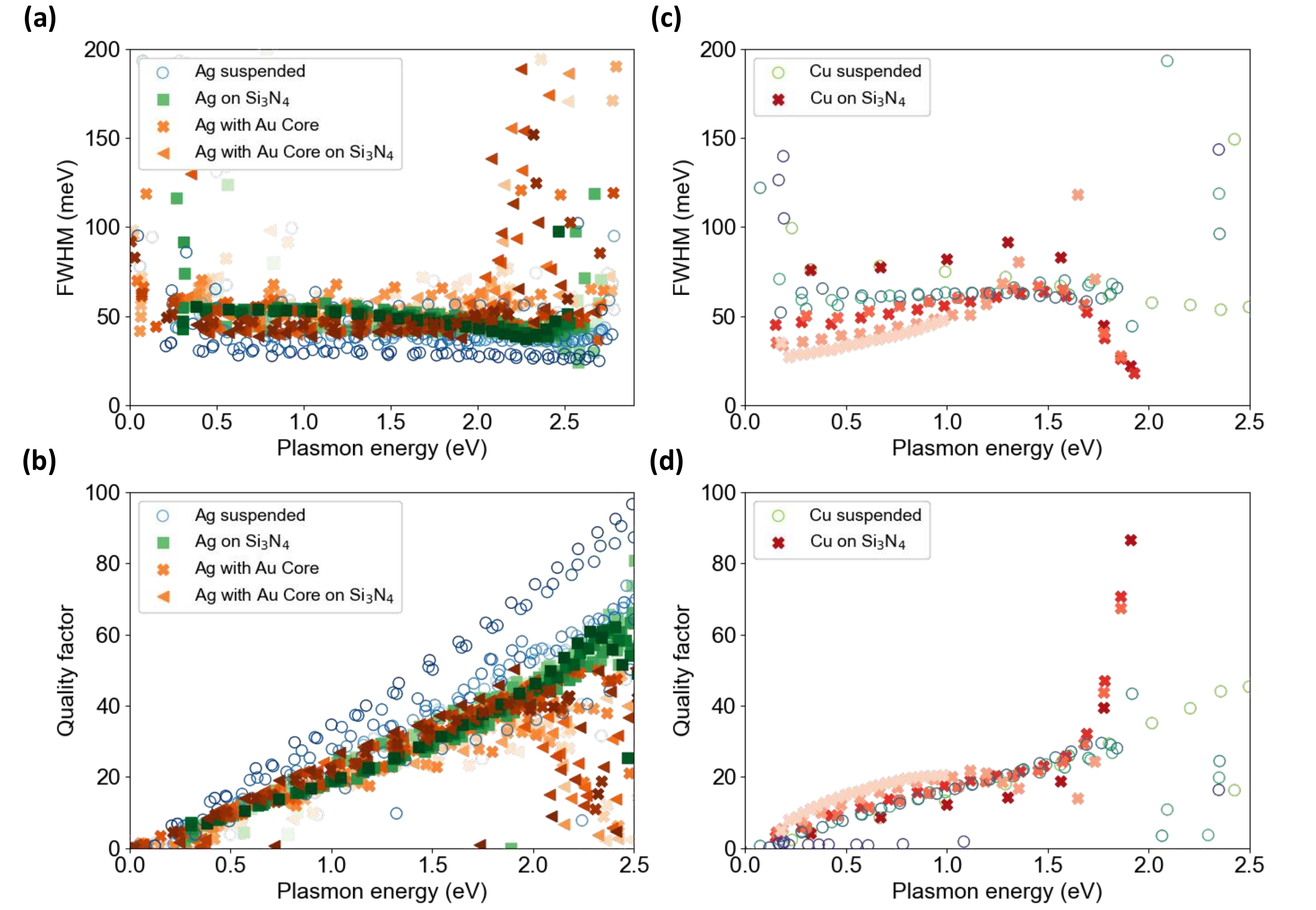}
\caption{Finite-difference time-domain (FDTD) simulations of the spectral width and quality factor $Q$ of Cu and Ag NWs. We consider (a,b) Ag and (c,d) Cu NWs in two different configurations as indicated in the legends: self-standing or supported on a 15\,nm Si$_3$N$_4$ membrane. For Ag NWs, we show calculations with and without the presence of a Au core at the center (220\,nm length, 32\,nm diameter).}
\label{FigS5}
\end{figure*}

\begin{figure*}[b]
\centering
\includegraphics[width=0.5\textwidth]{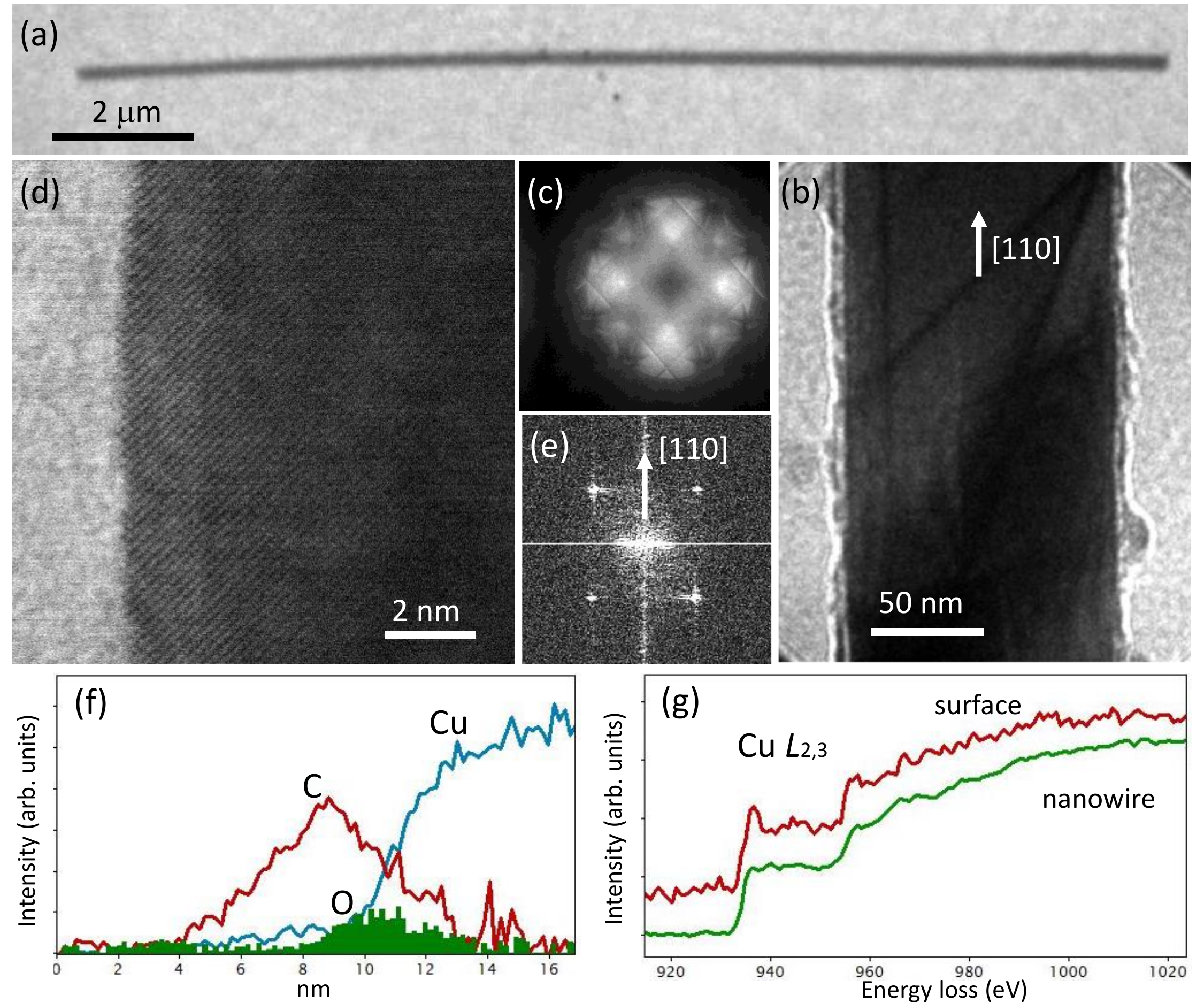}
\caption{Composition and crystallography of Cu nanowires (NWs). (a) Scanning-transmission-electron-microscope bright-field (STEM-BF) image of a long Cu NW. (b) Detail of the structure, revealing a $\sim120\,$nm diameter and a $[110]$ crystallographic orientation along the NW axis. (c) Converged beam electron diffraction (CBED, 30\,mrad convergence) taken near the center of the NW. (d) STEM-BF detail of the wire surface. The brightest area corresponds to a slightly oxidized region. (e)\,Fourier transform of (d), confirming the orientation of NW growth. (f) STEM-EELS compositional profile. The edge intensities are renormalized to the cross sections of the C-K, Cu-L, and O-K lines. The oxygen profile shown in green reveals that its content is weak and localized at the surface. (g) Cu $L_{2,3}$ edge EELS spectra acquired with the electron beam focused either on the Cu NW or 0.5\,nm away from the surface. We conclude that only a small amount of oxygen is observed several nanometers away from the surface, both when performing elemental EELS or when examining the change in the EELS Cu $L_{2,3}$ fine structure. This figure also shows a very limited degree of oxidation near the surface.}
\label{FigS1}
\end{figure*}

\begin{figure*}
\centering
\includegraphics[width=1.0\textwidth]{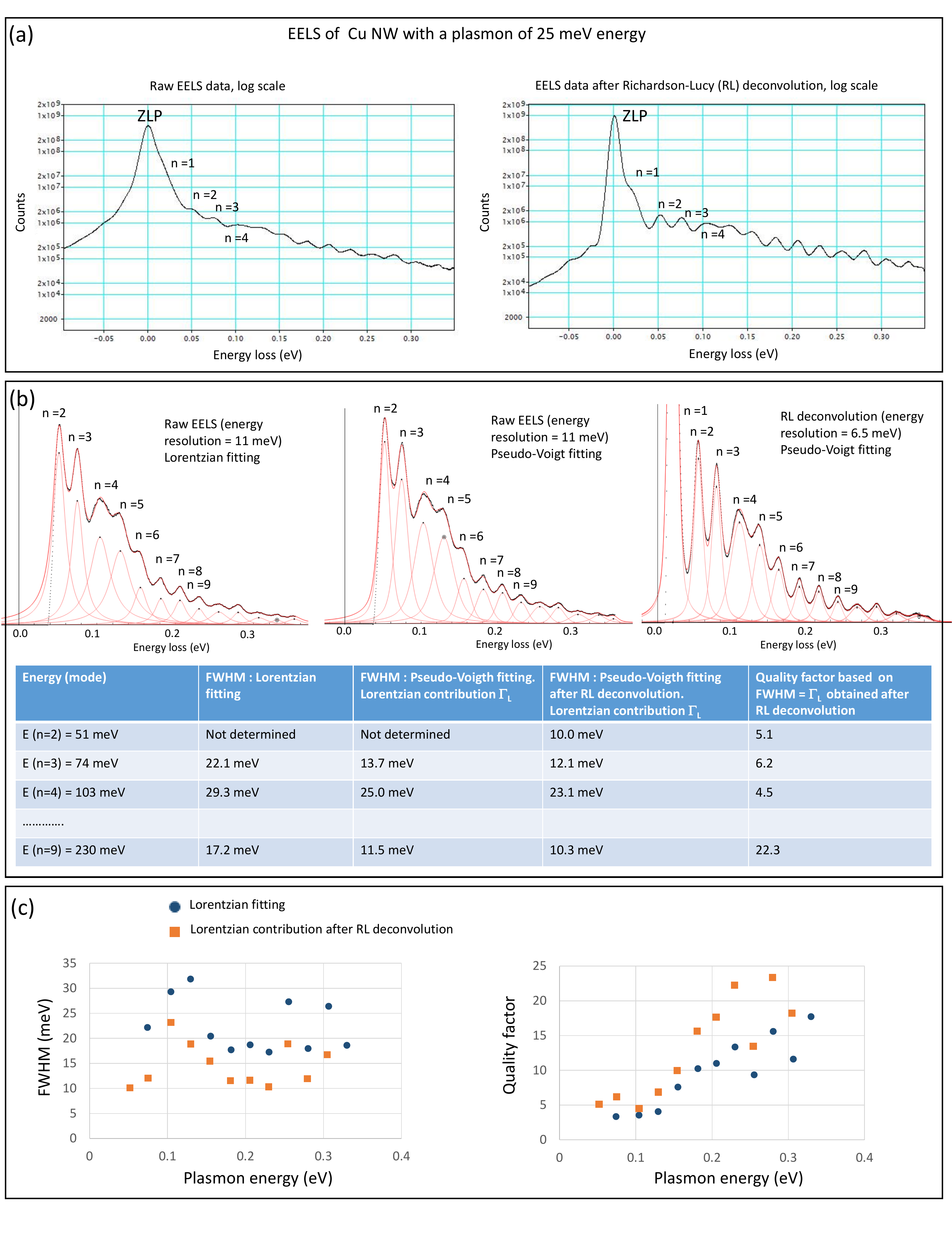}
\caption{Details of plasmon width determination. I. (a) Raw (left) and deconvoluted (right) EELS spectra of a Cu NW hosting a 25\,meV plasmon. (b) Examples of mode fitting. (c) FWHM and quality factor $Q$ obtained from a Lorentzian fitting of the raw data and from an estimate of the natural broadening (Voigt fitting and ZLP deconvolution).}
\label{FigS2}
\end{figure*}

\begin{figure*}
\centering
\includegraphics[width=0.6\textwidth]{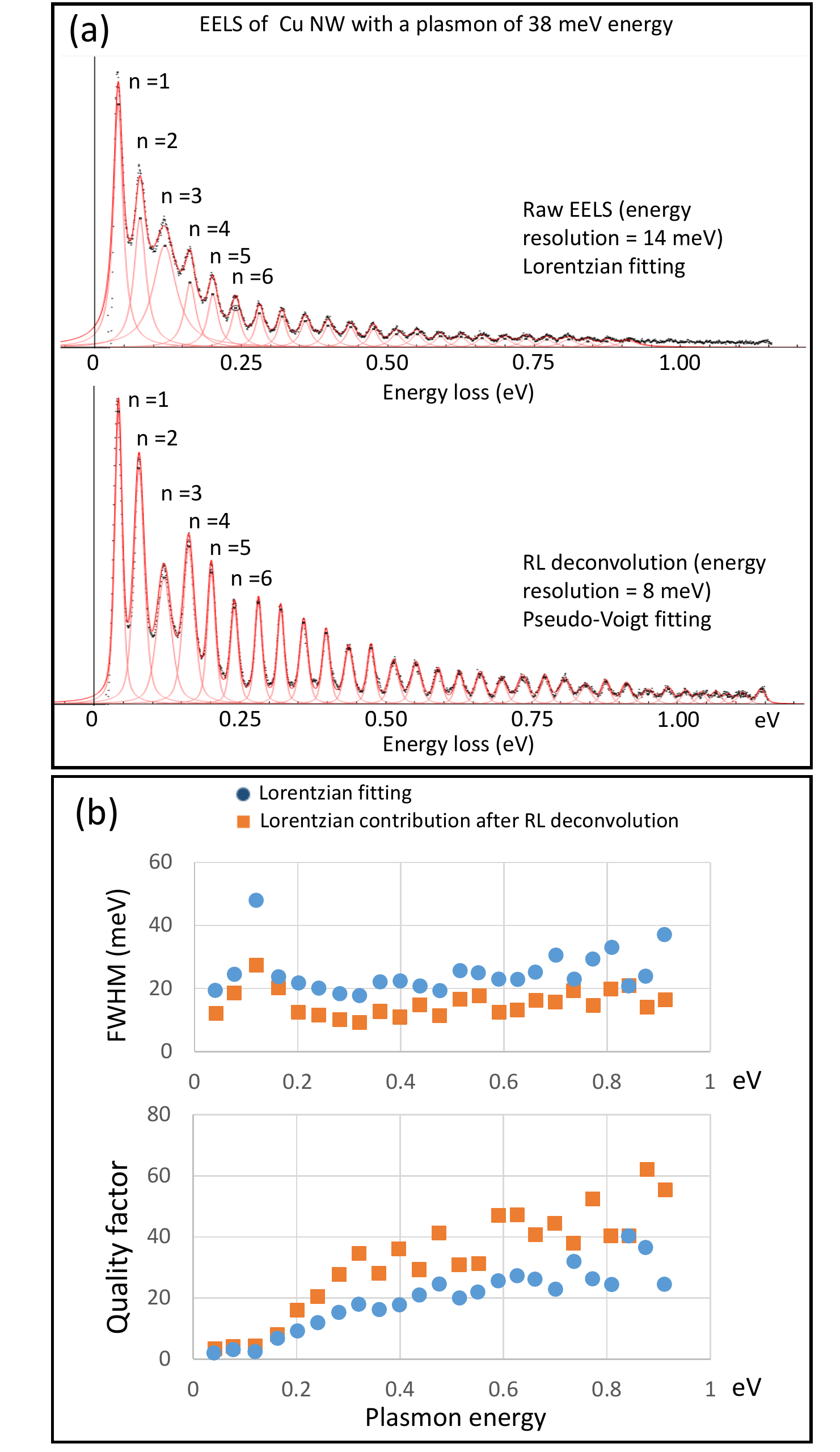}
\caption{Details of plasmon width determination. II. (a) Raw EELS spectra of a Cu NW hosting a 38\,meV plasmon. (b) RL deconvolution and pseudo-Voigt fitting. (c) FWHM and quality factor $Q$ obtained from this analysis.}
\label{FigS3}
\end{figure*}

\begin{figure*}
\centering
\includegraphics[width=0.6\textwidth]{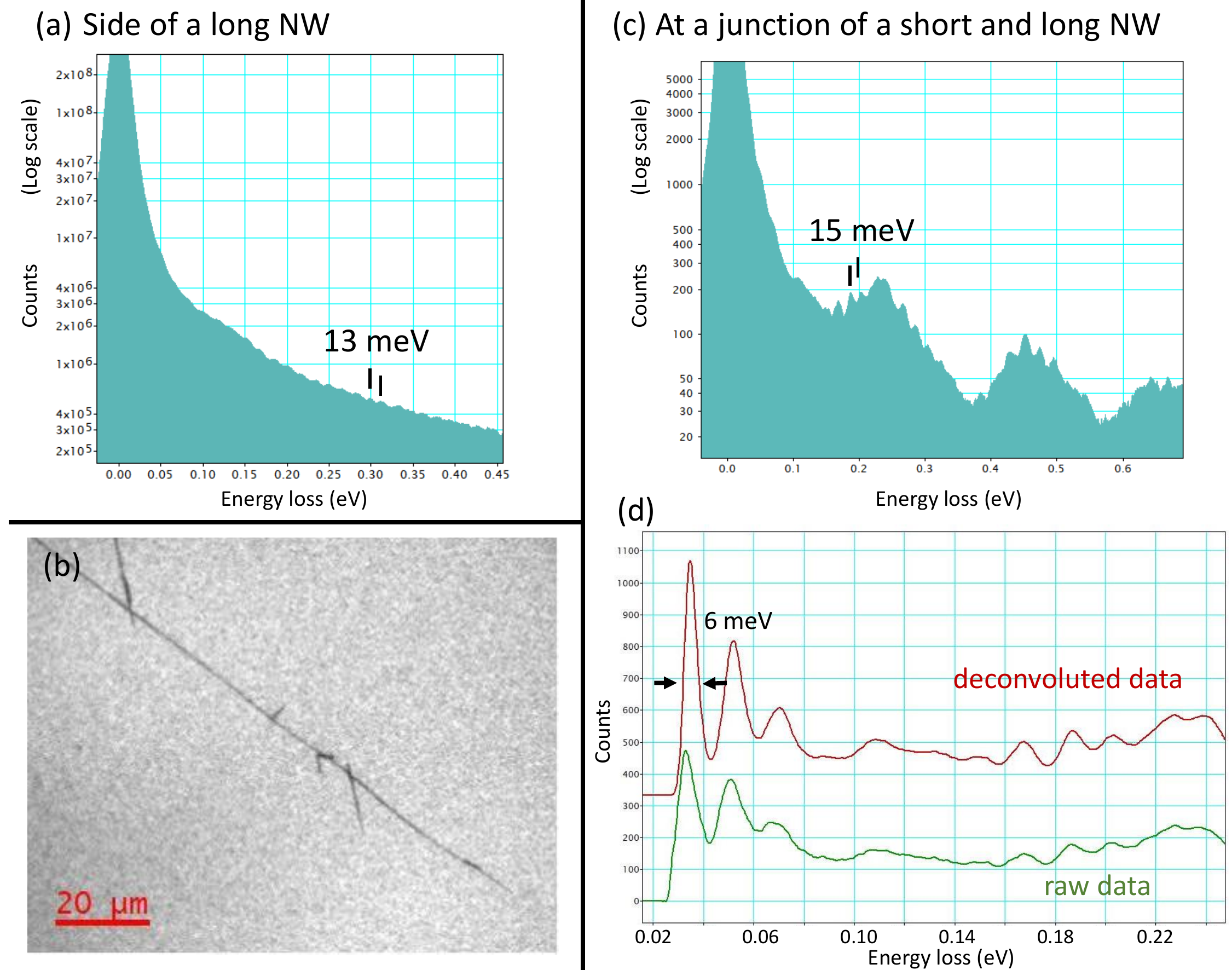}
\caption{Details of plasmon width determination. III. (a) EELS spectrum measured at a random position near the side wall of a long ($>100\,\mu$m) Cu NW. Plasmonic resonances are observed with an energy spacing $\sim13\,$meV. (b) BF image of interconnected long ($>100\,\mu$m) and shorter ($<10\,\mu$m) Cu NWs. (c) EELS spectra measured at a junction of long and short NWs. Plasmonic resonances with an intensity profile typical of beating frequencies are observed. (d) Raw ($\sim12\,$meV energy resolution) and deconvoluted ($\sim5\,$meV energy resolution) spectra after ZLP removal of the spectrum in (c). Some of the plasmonic features exhibit a FWHM well below 10\,meV (e.g., 6\,meV), even when collected with a ZLP of $\sim5\,$meV, indicating an extremely reduced damping ($<5\,$meV).}
\label{FigS4}
\end{figure*}

\end{document}